\documentclass[12pt]{article}

\usepackage{authblk}
\usepackage{graphicx}
\usepackage{amsmath}
\usepackage{mathtools}
\usepackage{braket}
\usepackage{amssymb}
\usepackage{geometry}
\usepackage{sectsty}
\usepackage[font=small,labelfont=bf]{caption}

\providecommand{\keywords}[1]{\small \textbf{Keywords: }#1}

\def\hh{\hspace{0.5mm}}

\def\dd{\mbox{d}}
\def\ddd{\scriptsize{\mbox{d}}}

\allsectionsfont{\sffamily}

\title{\sffamily {\bf Time Series Path Integral Expansions for Stochastic Processes}}

\author{Chris D Greenman}

\affil{\small{School of Computing Sciences, University of East Anglia, Norwich,
NR4 7TJ, United Kingdom. C.Greenman@uea.ac.uk}}

\begin{document}

\maketitle

%%%%%%%%%%%%%%%%%%%%%%%%%%%%%%%%%%%%%%%%%%%%%%%%%%%%%%%%%%%%%%%%%%%%%%%%%%%
%%%%%%%%%%%%%%%%%%%%%%%%%%%%%%%%%%%%%%%%%%%%%%%%%%%%%%%%%%%%%%%%%%%%%%%%%%%

\begin{abstract}
A form of time series path integral expansion is provided that enables both analytic and numerical temporal effect calculations for a range of stochastic processes. Birth-death processes with linear rates are analysed via coherent state Doi-Peliti techniques. The $\mathfrak{su}(1,1)$ Lie algebra is utilised to capture quadratic rate birth-death processes. The techniques are also adapted to diffusion processes. All methods rely on finding a suitable reproducing kernel associated with the underlying algebra to perform the expansion. The resulting series differ from those found in standard Dyson time series field theory techniques. 
\end{abstract}


%%%%%%%%%%%%%%%%%%%%%%%%%%%%%%%%%%%%%%%%%%%%%%%%%%%%%%%%%%%%%%%%%%%%%%%%%%%
%%%%%%%%%%%%%%%%%%%%%%%%%%%%%%%%%%%%%%%%%%%%%%%%%%%%%%%%%%%%%%%%%%%%%%%%%%%

\section{Introduction}
\label{I}

Deriving the time dependent behaviour for stochastic processes is a primary goal, irrespective of the type of process involved, and is the focus of interest presented herein. The processes considered in this work are continuous in time and may be discrete or continuous in state. Specifically, time dependent birth-death and diffusion processes are analysed.

For many models the analytic difficulty of obtaining short term temporal behaviour means that asymptotic techniques are studied. The long term behaviour and steady state conditions are of interest in their own right, and relevant techniques include renormalisation \cite{Tauber2014, Tauber2005} and system size expansion methods \cite{VanKampen1992}, to name a couple. However, it is short term behaviour considered in this work.

The classic discrete state processes are birth-death processes, and methods to obtain the population size distributions have been extensively studied. Solving the master equation for linear rates was achieved by Feller \cite{Feller1939} via Bartlett's generating function methods \cite{Bartlett1978}, with extension to time dependent rates by Kendall \cite{Kendall1948}. A general framework was derived by Karlin and McGregor \cite{Karlin1957b, Karlin1957a}, which relies on finding a suitable set of orthogonal polynomials involving rates that can be population and time dependent. The difficulty with this method is finding an analytic form for the polynomial orthogonality weights. This has been achieved for many types of birth-death rate, including linear, quadratic and quartic \cite{Sasaki2009,Valent1996}, although including time dependence generally makes this approach difficult.

The classic continuous state processes are diffusion processes, which were initially developed to analyse Brownian motion \cite{Wiener1921}, later developing into Ito calculus techniques \cite{Klebaner2012}. The state probability distribution is described by the Fokker-Planck equation, which has enabled alternative methods such as semi-group approaches \cite{Engel2001} to analyse such systems. 

Both discrete and continuous state processes can also be modelled via path integral techniques. Although these were popularised by Feynman \cite{Feynman2010}, they were initially developed by Daniell \cite{Daniell1919} for functional integration and by Weiner \cite{Wiener1923} from work on Brownian motion. Approaches utilising path integrals include Onsager-Machlup functional \cite{Onsager1953b,Onsager1953a} and Martin-Siggia-Rose approaches \cite{Martin1973}, developed primarily for diffusion processes. More recently, the use of field theoretic methods for discrete state processes was developed by Doi \cite{Doi1976a,Doi1976b} (for molecular reaction systems), and adapted to birth-death processes via path integral techniques by Peliti \cite{Peliti1985}. For further results on path integrals, see \cite{Kleinert2009, Wio2013}.

The use of field theory in birth-death processes has seen a wealth of development. This includes renormalization techniques that can be used to investigate asymptotic properties of systems and perturbation techniques to evaluate some path integrals of interest \cite{Tauber2014}. These expansion techniques are usually of the Feynman variety, where non quadratic terms in the associated action are expanded and generating functional techniques (or similar) are then applied. Such methods have also seen applications involving glass phase transition \cite{Garrahan2007}, branching random walks \cite{Cardy1998}, phylogeny \cite{Jarvis2005}, age structured systems \cite{Greenman2016, Greenman2017, Greenman2015}, neural network fluctuations \cite{Buice2007}, exclusion processes \cite{Greenman2018, Schulz2005, Schulz1996}, preditor prey systems \cite{Dobramysl2018}, stochastic duality \cite{Greenman2018, Ohkubo2013}, knot diagram dynamics \cite{Rohwer2015} and algebraic probability \cite{Ohkubo2013b}, to name a few. 

Here we take a different approach and evaluate the path integrals via time series expansion. Such an approach can be commonly found in quantum mechanics under the guise of Dyson series \cite{Peskin2018, Kleinert2009}. This normally involves using an interaction framework (bridging Heisenberg and Schrodinger representations) where the Hamiltonian is split into a standard and interacting component. The approach adopted here does not do this, rather it implements an exact calculation based on the entire Hamiltonian (or more specifically the Liouvillian, a designation for the Hamiltonian operator when applied to stochastic processes) and utilises a reproducing kernel framework to extract time series information.

This relies on an underlying algebra to work with. Classically, this is the Doi-Peliti framework derived from the ladder operators frequently seen with quantum harmonic oscillators, which can be used to model birth-death processes with linear rates, and can also be adapted to diffusion systems. Although these algebras can model birth-death processes with quadratic rates, this generally results in quartic terms in the action resulting in perturbation techniques. Instead we introduce an alternative system based on the $\mathfrak{su}(1,1)$ Lie algebra, which can more naturally deal with these systems. 

The paper is structured as follows. In \S\ref{LBDP} the expansion technique for linear rate birth-death processes via a standard Doi-Peliti structure are introduced. In \S\ref{QBDR} quadratic rate birth-death processes via a novel Doi-Peliti like structure are then considered, after which \S\ref{DE} adapts the techniques for diffusion processes. Finally, conclusions in \S\ref{ConcS} complete the manuscript.

%%%%%%%%%%%%%%%%%%%%%%%%%%%%%%%%%%%%%%%%%%%%%%%%%%%%%%%%%%%%%%%%%%%%%%%%%%%
%%%%%%%%%%%%%%%%%%%%%%%%%%%%%%%%%%%%%%%%%%%%%%%%%%%%%%%%%%%%%%%%%%%%%%%%%%%

\section{Linear Birth-Death Processes}
\label{LBDP}

This section examines birth-death processes with time dependent rates linear in population size. The next subsection describes a pedagogic model of interest and provides numerical results exemplifying the methods. Then in turn the algebra, the relevant path integral, the associated reproducing kernel, and expansion methods are described, finishing with a description of some of the difficulties seen when trying to adapt this method to Dyson expansions.

%%%%%%%%%%%%%%%%%%%%%%%%%%%%%%%%%%%%%%%%%%%%%%%%%%%%%%%%%%%%%%%%%%%%%%%%%%%

\subsection{A Spontaneous Annihilation-Immigration Process}
\label{ASAAP}

In order to highlight the methods, we consider the following process,
\begin{eqnarray}
A+A & \overset{\alpha_n(t)}{\longrightarrow} & \phi, \nonumber\\
\phi & \overset{\beta_n(t)}{\longrightarrow} & A.
\label{BDModel}
\end{eqnarray}
Thus we have a time dependent annihilation (death) and immigration (birth) process. Here the rates are given by $\alpha_n(t) = \alpha(t)n(n-1)$ and $\beta_n(t) = \beta(t)n$. Then we have linear population size dependence (giving quadratic dependence for the coming together of two annihilating particles) and time dependent functions $\alpha(t)$ and $\beta(t)$. We also assume the initial population size is Poisson distributed with mean value $w$. Analysing these kind of annihilation systems directly with classical methods is complicated by a moment closure problem. Specifically, using the master equation to get a dynamic equation for the first moment (mean) population size implicates the second moment, which itself requires a further dynamic equation, subsequently leading to a cascade of equations. This results in a BBGKY like hierarchy \cite{Bogoliubov1946,Born1949,Kirkwood1946,Kirkwood1947,Yvon1935}, making the evaluation of moments difficult, and the need for alternative methods is desirable.

\begin{figure}[t!]
\centering
\includegraphics[width=.8\linewidth]{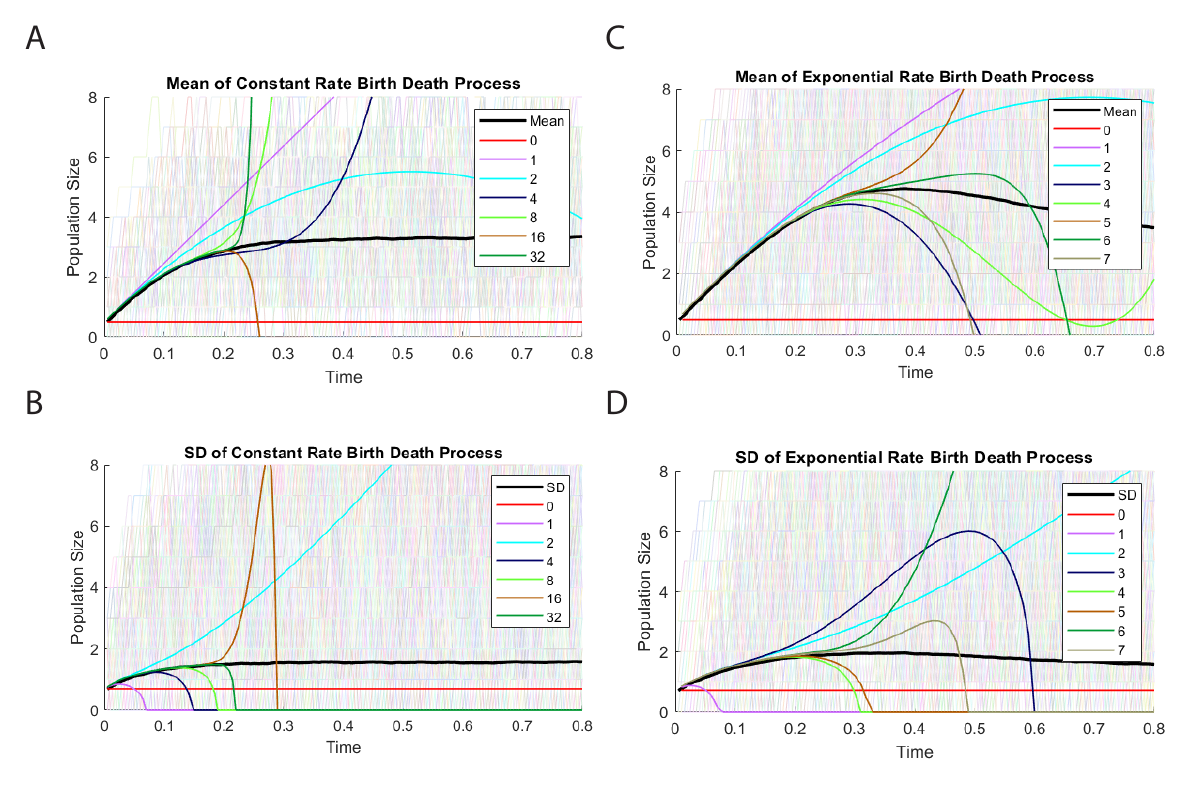}
\caption{Plots of population size mean and standard deviation for the process in Eq. \ref{BDModel}. The background of 5000 simulations is provided, along with the sample means (A,C) and standard deviations (B,D). The time series approximations using up to $32$ terms for constant rates $\alpha(t) = 1, \beta(t)=20$ in (A,B) is provided. The case of variable rates $\alpha(t) = 1-e^{-t}$ and $\beta(t)=20e^{-t}$ using up to $8$ time series terms are given in (C,D) (see text for further details). In all cases the initial population size had Poisson parameter $w=0.5$.}
\label{BDFig}
\end{figure}

In Fig. \ref{BDFig} we can see the results of simulating the process of Eq. \ref{BDModel}, along with the sample mean and standard deviation. This is done for the time independent case (Fig. \ref{BDFig}(A,B)), where the population size will eventually reach a non-trivial equilibrium. It is also done for the time dependent case (Fig. \ref{BDFig}(C,D)), with rates such that the initial growth will ultimately lose out to an increasing rate death process, resulting in extinction or the survival of a lone particle which cannot pairwise annihilate. The time series expansions (derived below) are also provided, showing a good fit for an initial time period. For the case of pure annihilation ($\beta=0$) or pure immigration ($\alpha=0$), closed form solutions can also be obtained (see below). The next few subsections describe the associated path integral machinery and expansion methods in more detail. 

%%%%%%%%%%%%%%%%%%%%%%%%%%%%%%%%%%%%%%%%%%%%%%%%%%%%%%%%%%%%%%%%%%%%%%%%%%%

\subsection{Ladder Operators}
\label{LO}

The algebra underlying the path integral for linear birth-death processes is the relatively standard ladder operators of quantum harmonic oscillators, so the overview is concise. However, some details of the underlying reproducing kernel are not normally discussed and are needed for the time series expansion, which will later be discussed in a bit more detail.

We have annihilation, creation and number operators $a$, $a^\dag$ and $N$, that obey the usual commutation relations,
\begin{eqnarray}
[a,a^\dag] & = & I,\nonumber\\
{[}N,a] & = & -a, \nonumber\\
{[}N,a^\dag] & = & a^\dag.
\label{Lie1}
\end{eqnarray}
Their action is defined on the orthogonal basis $\ket{n},n\in\{0,1,2\dots\}$, normalised via $\braket{m|n} = \delta_{mn}m!$, as follows,
\begin{eqnarray}
a\ket{n} & = & n\ket{n-1},\nonumber\\
a^\dag\ket{n} & = & \ket{n+1},\nonumber\\
N\ket{n} & = & n\ket{n}.
\end{eqnarray}

Now, if $p_n(t)$ represents the probability of population size $n$ at time $t$, we have state vector $\ket{\Phi_t} = \sum_n p_n(t)\ket{n}$ which satisfies evolution equation,
\begin{equation}
\frac{\partial}{\partial t} \ket{\Phi_t} = L\ket{\Phi_t},
\label{DynEq}
\end{equation}
for a suitable Liouvillian operator $L(a^\dag,a;t)$. For the model above, the associated Liovillian is given by \cite{Doi1976a, Doi1976b, Peliti1985},
\begin{equation}
L = \alpha(t)(a^2-(a^\dag)^2a^2)+\beta(t)(a^\dag-1).
\label{Liouv1}
\end{equation}

The last components needed are the coherent states. These are defined as $\ket{z}=e^{za^\dag}\ket{0}$ for possibly complex $z$, and have the following properties,
\begin{equation}
a\ket{z} = z\ket{z}, \hspace{5mm}
a^\dag\ket{z} = \frac{\partial}{\partial z}\ket{z}, \hspace{5mm}
\braket{x|z} = e^{\bar{x}z},
\label{EvalProp}
\end{equation}
where the second equation is meant in the sense that $\braket{x|a^\dag|z}=\frac{\partial}{\partial z}\braket{x|z}$.

Next consider the path integral construction.

%%%%%%%%%%%%%%%%%%%%%%%%%%%%%%%%%%%%%%%%%%%%%%%%%%%%%%%%%%%%%%%%%%%%%%%%%%%

\subsection{Path Integral Construction}
\label{PIC}

Now, to construct a path integral for features of interest such as correlation functions and the moments given in Fig. \ref{BDFig} there are a few approaches. One can use the Bargman-Fock formalism adopted by Peliti \cite{Itzykson2012,Peliti1985} which convert states and operators to functions. Other approaches vary depending on how the resolution of the identity is selected. The approach taken in \cite{Greenman2017} utilises a form that will be used in \S\ref{DE}. However, the following identity shall be adopted, which is a standard form more often seen in quantum field theory, where,
\begin{equation}
I = \int \frac{\dd\Re(z)\hh\dd\Im(z)\hh}{\pi}e^{-\bar{z}z}\ket{z}\bra{z} \equiv \int \frac{\dd z^2}{\pi} \hh e^{-\bar{z}z}\ket{z}\bra{z}.
\label{ROI1}
\end{equation}

Now, to exemplify the expansion method, consider constructing a path integral for the population size generating function, $G(x,w;t) = \sum_n p_n(t)x^n = \braket{x|\overleftarrow{\mathcal{T}}e^{\int_0^t \ddd s\hh L(a^\dag,a;s)}|\Phi_0}$. We assume that the initial population is Poisson distributed with mean value $w$, so $\ket{\Phi_0}= e^{-w}\ket{w}$. The Liouvillian $L(a^\dag,a;s)$ can be time dependent, so we have the time ordering operator $\overleftarrow{\mathcal{T}}$. The expansion method will utilise the path integral in an exact calculation, meaning the standard process of taking the continuum limit after time slicing to give an action does not suit our purposes. We take one step back, and find, using $n$ time slices (with $n\Delta = t$ and $t_k = k\Delta$) that,
\begin{align}
& e^w\braket{x|\overleftarrow{\mathcal{T}}e^{\int_0^t\ddd s \hh L(a^\dag,a;s)}|\Phi_0} =  
\int \prod_{k=0}^{n}\frac{\dd z_k^2}{\pi}\braket{x|z_n}\overleftarrow{\mathcal{T}}
\prod_{k=1}^n\braket{z_k|e^{\Delta L(a^\dag,a;t_k)}|z_{k-1}}
\braket{z_0|w}\nonumber\\
& = \int \prod_{k=0}^{n}\frac{\dd z_k^2}{\pi}
\exp\left\{\bar{x}z_n-\sum_{k=0}^n\bar{z}_kz_k + \sum_{k=1}^n\bar{z}_kz_{k-1}
+\Delta\sum_{k=1}^nL(\bar{z}_k,z_{k-1};t_k) + \bar{z_0}w \right\}\nonumber\\
& = \int \prod_{k=0}^{n}\frac{\dd z_k^2}{\pi}
\exp\left\{-\sum_{k=0}^n\bar{z}_kz_k + \sum_{k=0}^{n+1}\bar{z}_kz_{k-1}
 \right\}\cdot\nonumber\\
&\hspace{2cm}\prod_{k=1}^n\left(1+\Delta\left[\alpha(t_k)(z_{k-1}^2-\bar{z}_k^2z_{k-1}^2)+\beta(t_k)(\bar{z}_k-1)\right]\right).
\label{Expan1}
\end{align} 
Note that the weight notation has been simplified in the last line with the introduction of $z_{n+1} \equiv x$ and $z_{-1} \equiv w$. We are now left with an integration problem and time ordering has been resolved. The expansion method involves a path integral with the time slicing still in place. To proceed further, we need some reproducing kernel machinery, which we go through next.

%%%%%%%%%%%%%%%%%%%%%%%%%%%%%%%%%%%%%%%%%%%%%%%%%%%%%%%%%%%%%%%%%%%%%%%%%%%

\subsection{Reproducing Kernel Machinery}
\label{RKM}

Now a reproducing kernel $K(x,w)$ satisfies the property
\begin{equation}
K(x,w) = \int \dd \mu(z) K(x,z)K(z,w),
\end{equation}
for some suitable measure $\mu(z)$.

Now the main utility of the choice of resolution of identity in Eq. \ref{ROI1} is down to the following relationship, which can be found by pre and post multiplying $I$ by $\bra{x}$ and $\ket{w}$ (where $x$ and $w$ are for the moment treated as general and possibly complex).
\begin{equation}
e^{\bar{x}w} = \int \frac{\dd z^2}{\pi}  e^{\bar{x}z} e^{-\bar{z}z} e^{\bar{z}w}.
\label{KK1}
\end{equation} 
Thus we find that we have reproducing kernel $K(x,w) = e^{\bar{x}w}$ for the measure $\mu(z) = \frac{\ddd z^2}{\pi} e^{-\bar{z}z}$. Now, if we differentiate this form, we find the useful relationship,

\begin{equation}
\int \frac{\dd z^2}{\pi} e^{-\bar{z}z}\bar{z}^mz^n e^{\bar{x}z} e^{\bar{z}w}
= \partial_w^m\partial_{\bar{x}}^n \int \frac{\dd z^2}{\pi} e^{-\bar{z}z} e^{\bar{x}z} e^{\bar{z}w} = \partial_w^m\partial_{\bar{x}}^ne^{\bar{x}w},  
\end{equation}
where the shorthand $\partial_w \equiv \frac{\partial}{\partial w}$ is adopted. In particular, we find,
\begin{eqnarray}
\int \frac{\dd z^2}{\pi} e^{-\bar{z}z}z^n e^{\bar{x}z} e^{\bar{z}w}
& = & w^ne^{\bar{x}w},\nonumber\\  
\int \frac{\dd z^2}{\pi} e^{-\bar{z}z}\bar{z}^m e^{\bar{x}z} e^{\bar{z}w}
& = & \bar{x}^me^{\bar{x}w}.
\label{Sub}
\end{eqnarray}
Thus the $z$ and $\bar{z}$ powers get substituted with $w$ and $\bar{x}$, respectively. When both variables are present, the result is more complicated and we get,
\begin{eqnarray}
\int \frac{\dd z^2}{\pi} e^{-\bar{z}z}f(\bar{z},z)e^{\bar{x}z} e^{\bar{z}w} & = & f(\partial_w,w)e^{\bar{x}w},
\label{Recurr1a}\\
\int \frac{\dd z^2}{\pi} e^{-\bar{z}z}g(z,\bar{z})e^{\bar{x}z} e^{\bar{z}w} & = & g(\partial_{\bar{x}},\bar{x})e^{\bar{x}w}.
\label{Recurr1b}
\end{eqnarray}
Note in this last expression, we are treating the functions $f$ and $g$ as a power series in $\bar{z}$ and $z$, where the ordering is important. In $f$, $\bar{z}$ is left of $z$, meaning the $\partial_w$ terms are left of $w$, whereas in $g$ the order of $z$ and $\bar{z}$ is reversed. This somewhat akin to normal ordering of creation and annihilation operators, where the $a^\dag$ terms are left of $a$ terms. From these we can also recover the following expressions (where $f(z)$ and $g(\bar{z})$ are now taken to be functions of just one variable),
\begin{eqnarray}
\int \dd \mu(z)\frac{K(\bar{x},z)K(z,w)}{K(x,w)} f(z) & = & f(w),
\nonumber\\
\int \dd \mu(z)\frac{K(\bar{x},z)K(z,w)}{K(x,w)} g(\bar{z}) & = & g(\bar{x}),
\end{eqnarray}
and find that the kernels can be interpreted to play a role that Dirac delta functions typically provide. 

%%%%%%%%%%%%%%%%%%%%%%%%%%%%%%%%%%%%%%%%%%%%%%%%%%%%%%%%%%%%%%%%%%%%%%%%%%%

\subsection{Path Integral Expansion}
\label{PIE}

We now have what is needed to do the integration in Eq. \ref{Expan1}. Consider integrating with respect to $z_0$, which implicates factor $(1+\Delta L(z_0,\bar{z}_1;t_1))$ from the product, and three terms  $e^{-\bar{z_0}z_0}e^{\bar{z}_1z_0}e^{\bar{z}_0w}$ from the exponential weight. Note that we have now reversed the normal order of $z_0$ and $\bar{z}_1$ in the factor. The reason for this will be explained shortly, but is trivial to implement, as these are commutable complex numbers rather than operators. Then from Eq. \ref{Recurr1a} this integrates to $(1+\Delta L(w,\bar{z}_1;t_1))e^{\bar{z}_1w}$. Next we integrate factors $(1+\Delta L(w,\bar{z}_1;t_1))(1+\Delta L(z_1,\bar{z}_2;t_2))$ with weights $e^{-\bar{z_1}z_1}e^{\bar{z}_2z_1}e^{\bar{z}_1w}$ with respect to $z_1$, resulting in 
$(1+\Delta L(w,\partial_w;t_1))(1+\Delta L(w,\bar{z}_2;t_2))e^{\bar{z}_2w}$. Note again that the normal ordering is reversed, and also that second (later in time) factor is to the right. This form of time ordering allows $\bar{z}_1$ and $z_1$ to be normal ordered and Eq. \ref{Recurr1a} be used correctly. We repeat this process iteratively until we obtain,
\begin{equation}
e^wG(x,w;t) = \overrightarrow{\mathcal{T}}\lim_{n\rightarrow\infty}
\prod_{k=1}^n (1+\Delta L(w,\partial_w;t_k))e^{\bar{x}w}
= \overrightarrow{\mathcal{T}}e^{\int_0^t\ddd s\hh L(w,\partial_w;s)}e^{\bar{x}w}.
\end{equation}
Note the re-emergence of a time ordering operator $\overrightarrow{\mathcal{T}}$, although now time is ordered to the right. The final term obtained in the product will actually be $L(w,x;t_n)$. However, the term $x$ can be replaced by $\partial_w$ due to the action on rightmost factor $e^{\bar{x}w}$, resulting in the form above.

For most models of interest, the integration over time will be over rate parameters. For the model in Eq. \ref{Liouv1} we find,
\begin{equation}
e^wG(x,w;t) =\overrightarrow{\mathcal{T}} e^{A(t)(w^2-w^2\partial_w^2)+B(t)(\partial_w-1)}e^{\bar{x}w},
\label{Forw1}
\end{equation}
where $A(t) = \int_0^t \dd s \hh \alpha(s)$ and $B(t) = \int_0^t \dd s \hh \beta(s)$ are cumulative annihilation and immigration rates.

It is possible to do the integration in reversed time order. That is, integrate with respect to $z_n$ first and use Eq. \ref{Recurr1b} instead of \ref{Recurr1a}, resulting in an equation in $\bar{x}$ instead of $w$. The calculation is similar, although the difference between Eq.s \ref{Recurr1a} and \ref{Recurr1b} means the operator orders get reversed and we end up with,
\begin{equation}
e^wG(x,w;t) =\overleftarrow{\mathcal{T}} e^{A(t)(\partial_{\bar{x}}^2-\bar{x}^2\partial_{\bar{x}}^2)+B(t)(\bar{x}-1)}e^{\bar{x}w},
\label{Back1}
\end{equation}
where $\overleftarrow{\mathcal{T}}$ is the reverse time ordering operator.

Now differentiating either Eq. \ref{Forw1} or \ref{Back1} with respect to time produces the partial differential equation for the generating function, where we find (taking $x$ to be real),
\begin{equation}
G_t = \alpha(t)(1-x^2)G_{xx}
+\beta(t)(x-1)G.
\label{LPDEs}
\end{equation}
The initial condition is $G(x,w;0) = e^{w(x-1)}$. This equation could have been obtained directly by differentiating $e^wG = \braket{x|\overleftarrow{\mathcal{T}} e^{\int_0^t L(a^\dag,a;s) \ddd s}|w}$ with respect to time and letting $L$ act on $\bra{x}$. However, even with constant rates this PDE is not particularly trivial, and simplifying into canonical form \cite{Beals2016} is not that helpful. For example, solutions for the simpler case of pure annihilation ($\beta=0$, $\alpha(t) = \alpha$) are in series form and involve eigenvalue expansion techniques \cite{Mcquarrie1964}. 

%%%%%%%%%%%%%%%%%%%%%%%%%%%%%%%%%%%%%%%%%%%%%%%%%%%%%%%

\subsubsection{Pure Annihilation}

\begin{figure}[t!]
\centering
\includegraphics[width=.8\linewidth]{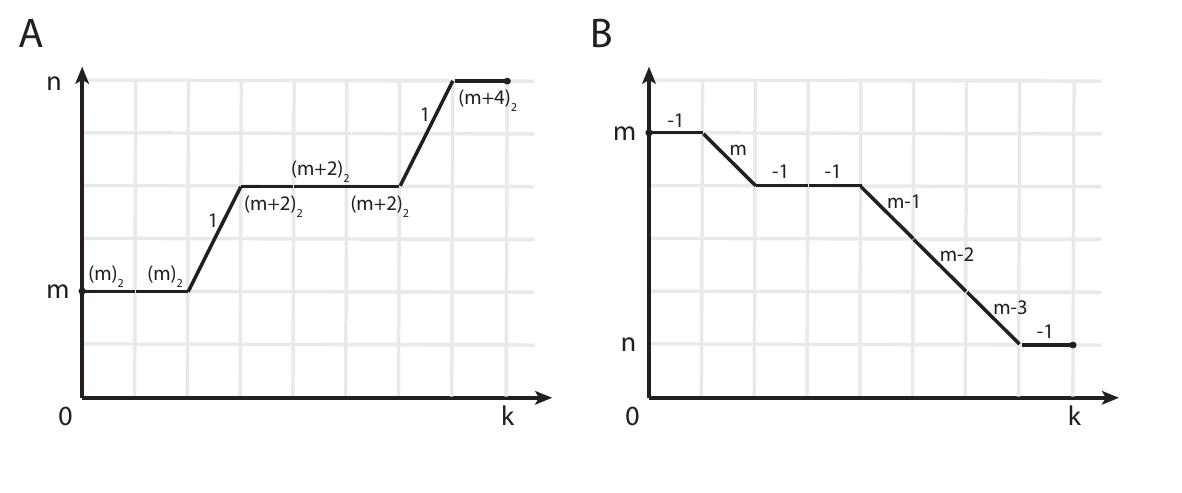}
\caption{Plots of paths associated with annihilation and immigration processes. Walks of $k$ steps going from height $m$ to $n$ are given. In (A), annihilation walks consist of double up steps or flat steps. In (B), immigration walks consist of horizontal or single down steps. In both cases weights associated with moves are given (see text for more details).}
\label{WalkFig}
\end{figure}

First then, consider the simpler case of pure annihilation ($\beta=0$), where we find that the time ordering operator plays no active role and can be ignored in Eq. \ref{Forw1}. Expanding in series we find that,
\begin{equation}
e^wG(x,w;t) = \sum_{k=0}^\infty \frac{(A(t))^k}{k!}(w^2-w^2\partial_w^2)^k\sum_{m=0}^\infty\frac{(\bar{x}w)^m}{m!}.
\end{equation}
Now we have the action of two operators to consider. In summary,
\begin{eqnarray}
w^2: w^m & \longrightarrow & w^{m+2}, \nonumber \\
-w^2\partial_w^2: w^m & \longrightarrow & -(m)_2w^m ,
\end{eqnarray}
where we adopt the Pochhammer notation $(m)_2 = m(m-1)$. Thus either the power of $w$ is increased by two with coefficient unity, or the power is unchanged and we pick up the factor $-(m)_2$. Now if we start with power $w^m$ and operate $k$ times this can be viewed as a discrete walk involving these two classes of step as seen in Fig \ref{WalkFig}(A). Any walk that starts from height $m$ and ends at height $n=m+2r$ in $k \ge r$ steps must use operator $w^2$, $r$ times, and operator $-w^2\partial^2$, $k-r=k-\frac{1}{2}(n-m)$ times. Now, each each distinct path connecting these points corresponds to a different order of operators, so it remains to combine the factors of the form $-(m)_2$ that emerge from the different paths. 

To this end, define $X_{k,n}$ as the product of coefficients $1$ and $-(m)_2$, summed across all paths of length $k$ that go from height $m$ to $n$ with $r = \frac{1}{2}(n-m)$ increasing steps each of size two. Now, we can form a recurrence by conditioning over a single step to find that;
\begin{eqnarray}
X_{k,n} & = & -(n)_2X_{k-1,n} + X_{k-1,n-2}, \nonumber \\
X_{0,n} & = & \delta_{nm}.
\end{eqnarray}

One can then attempt generating function approaches to solve the recurrence. However, by inspection, we note that if we take $k-r$ values (with repetition allowed) from the set $\{-(m)_2,-(m+2)_2,\dots,-(n)_2\}$ and order them, there is a corresponding path (provided $n\ge m$ have the same parity), and so we sum over the choices (so sum over size $k-r$ subsets with repetition allowed) to give,
\begin{equation}
X_{k,n} = \sum_{\{\underline\eta:\sum_{i=0}^r \eta_i =  k-r\}}(m)_2^{\eta_0}(m+2)_2^{\eta_1}\dots(m+2r)_2^{\eta_r}(-1)^{k-r}.
\label{Xpaths}
\end{equation}
These are encapsulated by the terms found from expanding $(-(m)_2-(m+2)_2-\dots-(m+2r)_2)^{k-r}$ and replacing the multinomial coefficients with unity. Note that $X_{k,n}$ is a function of $m$; the dependence is suppressed for ease of presentation. We now have a sum over paths. Take for example the term $(-(m)_2)^{k-r}$ from the sum $X_{k,n}$, which corresponds to the path with $k-r$ horizontal steps followed by $r$ increases, in turn corresponding to the operator product with order $(w^2)^r(-w^2\partial_w^2)^{k-r}$. Thus we can write the solution as follows, taking the form of a weighted inner product of two exponential series,
\begin{equation}
G(x,z;t) = e^{-w}\sum_{k=0}^\infty \frac{A(t)^k}{k!}\sum_{m=0}^\infty\frac{x^m}{m!}
\sum_{\substack{\{n:m\le n \le m+2k\\n\equiv m \mod(2)\}}}X_{k,n}w^n.
\end{equation}

%%%%%%%%%%%%%%%%%%%%%%%%%%%%%%%%%%%%%%%%%%%%%%%%%%%%%

\subsubsection{Pure Immigration}

The solution for the case of $\alpha=0$ (pure immigration) is similar, except now we have decreasing steps of size one, rather than increasing steps of size two, to consider. From Eq. \ref{Forw1}, the operator actions of interest are now,
\begin{eqnarray}
\partial_w: w^m & \longrightarrow & m w^{m-1}, \nonumber \\
-1: w^m & \longrightarrow & -w^m.
\end{eqnarray}
Then again consider paths of $k$ steps going from height $m$ to $n$. All the horizontal steps have factor $-1$. The first down step has factor $m$, and the next down step has factor $m-1$, irrespective of whether horizontal steps have occurred meanwhile. Then when $r$ down steps take place, we get factor $(m)_r$. Let $X_{k,n}$ sum these factors over paths, which follows a simpler recursion of the form $X_{k,n} = -X_{k-1,n}+(n+1)X_{k-1,n+1}$. Note that requiring $n$ to be non-negative limits the number of down steps. Now, it doesn't matter where all the $m-n$ down steps are positioned relative to the horizontal steps, the same factors occur, giving $(m)_{m-n} = \frac{m!}{n!}$ (see Fig. \ref{WalkFig}(B)). Also, the number of ways to position the $k-m+n$ horizontal steps is $k \choose {k-m+n}$. Thus we obtain,
\begin{equation}
G(x,z;t) = e^{-w}\sum_{k=0}^\infty \frac{B(t)^k}{k!}
\sum_{m=0}^\infty\frac{x^m}{m!} 
\sum_{n=\max\{0,m-k\}}^{m}X_{k,n}w^n,
\end{equation}
where,
\begin{equation}
X_{k,n} = (m)_{m-n}{k\choose k-m+n}(-1)^{k-m+n}.
\end{equation}

Now, if the summation over $n$ is moved to the left, after a little algebra, we find $G(x,w;t) = e^{(x-1)(B+w)}$. The simplicity of this result is reflected in the fact that for this example, the result can be more easily obtained from the first order PDE derived from Eq. \ref{LPDEs}, where $G_t = \beta (x-1)G$, and even more trivially from the generating function expression in Eq. \ref{Back1}, or from the observation that the arrivals are Poisson with rate found from adding the initial rate $w$ to the cumulative arrival rate $B(t)$.

%%%%%%%%%%%%%%%%%%%%%%%%%%%%%%%%%%%%%%%%%%%%%%%%%%%%%

\subsubsection{Annihilation and Immigration with Constant Rates}
 
The case of mixed processes involves time ordering and is more involved. For the time independent case of constant rates ($\alpha(t)=\alpha$ and $\beta(t)=\beta$), the problem can be approached in a similar manner to above, although the path walking element now has up, down and horizontal steps, going from $m$ to $n$. We now have two rates involved, meaning we have a recurrence of the following form,
\begin{eqnarray}
X_{k,n} & = & \alpha X_{k-1,n-2}-\alpha(n)_2X_{k-1,n}
+\beta(n+1)X_{k-1,n+1}-\beta X_{k-1,n}, \nonumber \\
X_{0,n} & = & \delta_{n,m}.
\end{eqnarray}
Then we similarly find a solution of the form,
\begin{equation}
G(x,w;t) = e^{-w}\sum_{k=0}^\infty \frac{t^k}{k!}\sum_{m=0}^\infty\frac{x^m}{m!}
\sum_{n=\max\{0,m-k\}}^{m+2k}X_{k,n}w^{n}.
\end{equation}
This simplifies slightly if we introduce $Y_{k,n} = \sum_{m=0}^\infty\frac{x^m}{m!}\frac{t^k}{k!}X_{k,n}w^n$, which has corresponding recurrence,
\begin{eqnarray}
Y_{k,n} & = & (\alpha w^2Y_{k-1,n-2}-\alpha t(n)_2Y_{k-1,n}
+\beta w^{-1}t(n+1)Y_{k-1,n+1}-\beta tY_{k-1,n})/k, \nonumber \\
Y_{0,n} & = & e^{(x-1)w},
\end{eqnarray}
along with generating function,
\begin{equation}
G(x,w;t) = \sum_{k=0}^\infty\sum_{n=0}^\infty Y_{k,n}.
\end{equation}

This later form was used in Fig. \ref{BDFig}(A,B), where the mean and standard deviation of population size  were extracted from the generating function in the standard way (e.g. $\mathbb{E}(N) = G_x(1,w)$), for $k \le 32$. Higher order approximations started to run into numerical issues and likely require more than the standard double precision that was utilised in Matlab.

%%%%%%%%%%%%%%%%%%%%%%%%%%%%%%%%%%%%%%%%%%%%%%%%%%%%%%%%%%%%%%%%%%%%%%%%%%%

\subsection{Annihilation and Immigration with Time Dependent Rates}

For time dependent mixed processes the time ordering plays an active role. Now, expanding Eq. \ref{Forw1}, we find generating function,
\begin{equation}
e^wG(x,w;t) = \overrightarrow{\mathcal{T}} \sum_{k=0}^\infty\frac{(A(t)\mathcal{A}+B(t)\mathcal{B})^k}{k!}e^{\overline{x}w}.
\end{equation} 
Here $\mathcal{A} = w^2-w^2\partial_w^2$ and $\mathcal{B} = \partial_w-1$ are operators. Note that although $\mathcal{A}$ and $\mathcal{B}$ are not time dependent, the time ordering operator $\overrightarrow{\mathcal{T}}$ is acting upon $A(t)\mathcal{A}$ and $B(t)\mathcal{B}$ and so they are time ordered.

If we focus on the expected population size, we then require,
\begin{equation}
G_x(1,w;t) = e^{-w}\overrightarrow{\mathcal{T}}\sum_{k=0}^\infty\frac{(A(t)\mathcal{A}+B(t)\mathcal{B})^k}{k!}we^{w}.
\label{EEE}
\end{equation}
Consider these term by term. The zeroth order term ($k=0$) can be simply read off as $w$, the initial mean of the population. The first order term ($k=1$) requires no time ordering and we obtain,
\begin{equation}
e^{-w}(w^2-w^2\partial_w^2)we^wA(t)
+\hh e^{-w}(\partial_w-1)we^wB(t) = -2w^2\int_0^t\dd s\hh\alpha(s)+\int_0^t\dd s\hh\beta(s).
\end{equation}

Second order and higher order terms require time ordering to be observed. For example, the second order term arising from operator product $\mathcal{B}\mathcal{A}$ would be,
\begin{equation}
e^{-w}\int_{0<s<s'<t}\dd s \hh \dd s' \beta(s)\alpha(s')(\partial_w-1)(w^2-w^2\partial_w^2)we^w = 
-4w\int_{0<s<s'<t}\beta(s)\alpha(s').
\end{equation}
The term corresponding to $\mathcal{A}\mathcal{B}$ comes out as zero because $(w^2-w^2\partial_w^2)(\partial_w-1)we^w$ vanishes. The full second order term is,
\begin{equation}
(4w^2+8w^3)\iint_{0<s_1<s_2<t}\dd s_1 \dd s_2 \alpha(s_1)\alpha(s_2) + 
(-4w)\iint_{0<s_1<s_2<t}\dd s_1 \dd s_2 \beta(s_1)\alpha(s_2).
\label{SecOrd}
\end{equation}

The breakdown up to third order terms is given in Table \ref{Coeffs}, where the ordered rates to be integrated, and the associated functions of $w$ are given. These features can be built up iteratively. The action of operators $\mathcal{A}$ and $\mathcal{B}$ can be expressed generally in the form,
\begin{eqnarray}
\mathcal{A} f(x)\cdot e^x & = & x^2(f_{xx}-2f_x)\cdot e^x\nonumber\\
\mathcal{B} f(x)\cdot e^x & = & f_x\cdot e^x.
\end{eqnarray}
Also, the coefficient of $\beta\alpha\alpha$ given in Table \ref{Coeffs} is simply obtained by differentiating the coefficient for $\alpha\alpha$. 

For the example in Fig. \ref{BDFig}(C,D) the approximations up to order eight are given, which took about three hours on a laptop. The rates in this example can be integrated as an algebraic recursion, allowing for a relatively simple implementation. This was initially done with the aid of Matlab's symbolic toolbox, although direct coding of the recursion was much quicker. Increasing the expansion's order eventually resulted in memory and run time issues (due to the exponential number of terms), rather than precision became a limiting factor.

\begin{table}[h!]
\centering
\begin{tabular}{||c|c|c||} 
\hline
Order ($k$) & Rates & $w$ Term \\
\hline\hline
$0$ & - & $w$ \\ 
\hline
$1$ & $\alpha$ & $-2w^2$  \\
    & $\beta$ & 1 \\
\hline
$2$ & $\alpha\alpha$ & $4w^2+8w^3$ \\
    & $\alpha\beta$ & $0$ \\
    & $\beta\alpha$ & $-4w$ \\
    & $\beta\beta$ & $0$ \\
\hline
\end{tabular} 
\quad
\begin{tabular}{||c|c|c||} 
\hline
Order ($k$) & Rates & $w$ Term \\
\hline\hline
$3$ & $\alpha\alpha\alpha$ & $-8w^2-64w^3-48w^4$ \\
    & $\alpha\alpha\beta$ & $0$ \\
    & $\alpha\beta\alpha$ & $8w^2$ \\
    & $\alpha\beta\beta$ & $0$ \\ 
    & $\beta\alpha\alpha$ & $8w+24w^2$ \\
    & $\beta\alpha\beta$ & $0$ \\
    & $\beta\beta\alpha$ & $-4$ \\
    & $\beta\beta\beta$ & $0$ \\ 
 \hline
\end{tabular}
\caption{Coefficients for time dependent annihilation immigration model. The order is the $k^\mathrm{th}$ term in series from Eq. \ref{EEE}, given up to $k=3$. The rates are the time ordered rates to be integrated from $0$ to $t$. The $w$ term indicates the weight of each integrated order rate.}
\label{Coeffs}
\end{table}

Some comments are warranted. To calculate the $r^\mathrm{th}$ factorial moment, we just replace $we^w$ with $w^re^w$ in Eq. \ref{EEE}. Note that the integrated rates remain identical irrespective of what moment is considered. The time component is also the hardest part of the calculation, which was done recursively via Matlab. For constant rates the calculation is trivial resulting in terms of the form $\frac{\alpha^m\beta^nt^{m+n}}{(m+n)!}$, although then the approach of the previous subsection becomes applicable. Note that the recursion for integration of the rates (reverse time) is in opposite order to the operator recursion (forward time) which need aligning when implemented. For the more general setting, using families of rate functions closed under the rate recurrence will avoid getting into intractable integration. The model used in Fig. \ref{BDFig}(C,D) belonged to the family of terms with the form $t^ne^{\alpha t}$ which are closed under the recursion making the calculation tractable, which will not be the case more generally. This has all been implemented by acting on the initial state variable $w$. However, the expansion could have been performed using the generating function variable $x$ with Eq. \ref{Back1}, although then the operators $\mathcal{A}$ and $\mathcal{B}$ act on $x$ as does the operator $\partial_x$ needed to extract moments from the generating function and some care is needed. Time ordering is also reversed in this case.

%%%%%%%%%%%%%%%%%%%%%%%%%%%%%%%%%%%%%%%%%%%%%%%%%%%%%%%%%%%%%%%%%%%%%%%%%%%

\subsection{Interaction Picture Perturbation}

It is natural to question what happens when the classical Dyson series approach of time dependent perturbation is adapted to the methods above. Dyson series perturbation uses the interaction picture, which assumes that the Liouvillian can be split up as $L = H + V(t)$, where $H$ is a well behaved part, and the remaining terms in $V$ treated perturbatively. Typically, time dependencies are contained in $V(t)$.

Then next we introduce the state (in the interaction picture) $\ket{\xi_t}_I = e^{-Ht}\ket{\Phi_t}$ (so that $\ket{\xi_0}_I=\ket{\Phi_0}$) and operator (in interaction picture) $V_I(t) = e^{-Ht}V(t)e^{Ht}$. This leads to the following conditions, with time ordering operator $\overleftarrow{\mathcal{T}}$,
\begin{eqnarray}
\frac{\partial}{\partial t}\ket{\xi_t}_I & = & V_I(t)\ket{\xi_t}_I,\nonumber\\
\ket{\xi_t}_I & = & \overleftarrow{\mathcal{T}}e^{\int_0^t\ddd s\hh V_I(s)}\ket{\xi_0}_I.
\end{eqnarray}

Then if we consider a feature of interest, such as the factorial moments, we find,
\begin{eqnarray}
\mathbb{E}((N_t)_m) & = & \braket{1|a^m|\Phi_t}
= \braket{1|a^me^{Ht}\sum_{m=0}^\infty \int_{0<s_1<\dots<s_m<t}V_I(s_m)\dots V_I(s_1)|\Phi_0}.
\label{DysExp}
\end{eqnarray}

Now to progress further we need to use resolutions of identity between operators. We do this for the example in Eq. \ref{BDModel}, where we use $H = \beta (a^\dag-1)$ and $V(t) = \alpha(t)(1-(a^\dag)^2)a^2$. We assume $\beta(t)=\beta$ is constant and $\alpha(t)$ is time dependent. Then consider the required terms. Firstly we find using Eq. \ref{EvalProp} that,
\begin{equation}
\braket{z|e^{Ht}|z'} = e^{\bar{z}z'}e^{\beta t(\bar{z}-1)}.
\end{equation}
From this we find (using the resolution of identity twice) that,
\begin{eqnarray}
\braket{z|V_i(s)|z'} & = & \iint \frac{\dd x^2}{\pi}\frac{\dd y^2}{\pi}e^{-\bar{x}x}
e^{-\bar{y}y}\braket{z|e^{-Ht}|x}
\braket{x|V(t)|y}\braket{y|e^{Ht}|z'}\nonumber\\
& = & \alpha(t)e^{-\beta t(\bar{z}-1)}e^{\bar{z}z'}(z')^2(1-\bar{z}^2).
\end{eqnarray}

Note that the time variable is not separable from the integration variables (cf. Eq. \ref{SecOrd}) and we are left with difficult integrations if we attempt to use this with Eq. \ref{DysExp}. Furthermore, most time dependent stochastic models of interest will have rates that either all depend upon time, or none do, making the extraction of a simpler component $H$ somewhat redundant.

%%%%%%%%%%%%%%%%%%%%%%%%%%%%%%%%%%%%%%%%%%%%%%%%%%%%%%%%%%%%%%%%%%%%%%%%%%%
%%%%%%%%%%%%%%%%%%%%%%%%%%%%%%%%%%%%%%%%%%%%%%%%%%%%%%%%%%%%%%%%%%%%%%%%%%%

\section{Quadratic Birth-Death Rates}
\label{QBDR}

Next birth-death processes with quadratic rates are considered. Such rates can be modelled with a standard Doi-Peliti approach. The resultant Liouvillian will involve quartic terms, which then can be treated with perturbative methods, including the expansion methods of above. However, there are alternative algebras to that in Eq. \ref{Lie1} which may be utilised, which we now demonstrate. This hinges on finding a suitable resolution of identity and associated reproducing kernel which is now explored.

%%%%%%%%%%%%%%%%%%%%%%%%%%%%%%%%%%%%%%%%%%%%%%%%%%%%%%%%%%%%%%%%%%%%%%%%%%%

\subsection{The Model}
\label{TM}

The following model will eventually be considered,

\begin{eqnarray}
A & \overset{\alpha_n(t)}{\longrightarrow} & A+A, \nonumber\\
A & \overset{\beta_n(t)}{\longrightarrow} & \phi.
\end{eqnarray}
That is, we have standard birth and death, with rates that are quadratic in population size. The overall rates $\alpha_n(t) = \alpha(t) (n^2+\nu n),$ and $\beta_n(t) = \beta(t) (n^2+\nu n)$ can be time dependent. The linear adjustment $\nu$ is a fixed positive real number which adds a bit of flexibility, but is fixed across time and both processes.

%%%%%%%%%%%%%%%%%%%%%%%%%%%%%%%%%%%%%%%%%%%%%%%%%%%%%%%%%%%%%%%%%%%%%%%%%%%

\subsection{The Algebra}
\label{TA}

First we consider a general framework that the above model will be applied to. We introduce an orthogonal set of states $\ket{n}, n \in \{0,1,\dots\}$, this time with normalisation $\braket{m|n} = \delta_{mn}\Gamma( n+1)\Gamma(n+\nu+1)$. It is assumed that $p_n(t)$ is the population size probability distribution for the system of interest. To look for an appropriate algebra for the system, the usual state vector is assumed,
\begin{equation}
\ket{\Phi_t} = \sum_{n=0}^\infty p_n(t) \ket{n}.
\end{equation}

Next annihilation, creation and number operators, $a$, $a^\dag$ and $N$, respectively, are introduced such that,
\begin{eqnarray}
a\ket{n} & = & (n^2+\nu n)\ket{n-1},\nonumber\\
a^\dag\ket{n} & = & \ket{n+1},\nonumber\\
N\ket{n}& = & n\ket{n}.
\end{eqnarray}
Then by letting commutators act on $\ket{n}$ we find the following non-trivial relations,
\begin{eqnarray}
[a,a^\dag] & = & 2N + \nu + 1 \equiv 2M ,\nonumber\\
{[}M,a] & = & -a,\nonumber\\
{[}M,a^\dag] & = & a^\dag.
\label{Lie2}
\end{eqnarray}
Thus we obtain the operator triple $M$, $a$ and $a^\dag$ which obey the commutation relations of semi-simple Lie algebra $\mathfrak{su}(1,1)$. Note that this three dimensional Lie algebra is not isomorphic to the standard one adopted from quantised harmonic oscillators from the previous section. More specifically, the Lie algebra $\mathfrak{g}$ of the quadruple $I$, $a$, $a^\dag$ and $N$ in Eq. \ref{Lie1} is solvable, with the upper series $\mathfrak{g}^{(3)}=0$ terminating after three commutations. Alternatively, if we factor out (in turn) the ideals $\{I\}$, $\{a,a^\dag\}$ and $\{N\}$, we can resolve $\mathfrak{g}$ as $\mathfrak{t}(1)\oplus_s\mathfrak{t}(2)\oplus_s\mathfrak{t}(1)$. In both cases we are using an infinite dimensional representation of the relevant Lie algebra \cite{Gilmore2012, Sattinger2013}.

Now with these relations the form of the dynamic equation is still that seen in Eq. \ref{DynEq}, except this time to model the system above we have a Liouvillian of the form,
\begin{equation}
L = \alpha(t)((a^\dag)^2a-a^\dag a)+\beta(t)(a-a^\dag a).
\end{equation}
Note that this is the standard Liouvillian for a birth-death process, except that now the rates are quadratic in population size rather than linear. This suggests we have a natural framework for the model. The rates $\alpha(t)$ and $\beta(t)$ are time dependent functions left in an unspecified form.

We again need the notion of a coherent state. These are now defined as,
\begin{equation}
\ket{z} = \sum_{m=0}^\infty\frac{(za^\dag)^m}{\Gamma(m+1)\Gamma(m+\nu+1)}\ket{0},
\end{equation}
which (after some manipulations using the commutation relations) have the following properties,
\begin{equation}
a\ket{z} = z\ket{z}, 
\hspace{3mm}
a^\dag\ket{z} = z^{-\nu}\partial_z z^{1+\nu}\partial_z \ket{z},
\hspace{3mm}
\braket{x|z} = \sum_{m=0}^\infty \frac{(\bar{x}z)^m}{\Gamma(m+1)\Gamma(m+\nu+1)} = \frac{I_{\nu}(2\sqrt{\bar{x}z})}{|\bar{x}z|^{\nu/2}} 
\equiv \hat{I}_v(\bar{x}z),
\label{Eig2}
\end{equation}
where $\partial_z \equiv \frac{\partial}{\partial z}$, and we have the modified Bessel function of the first kind in the last expression \cite{Bowman2012}. The unorthodox modified, modified Bessel function $\hat{I}_v$ is introduced purely to simplify expressions.

Finally we note that in this framework the coherent state $\bra{1}$ stills plays the same role as a sum over states. However, note that moments are extracted differently. We find, for example (using $N_t$ as the population size random variable, distinct from the number operator $N=\frac{1}{2}(aa^\dag-a^\dag a-I)$),
\begin{eqnarray}
\mathbb{E}(N_t^m) & = & \braket{1|N^m|\Phi_t},\nonumber\\
\mathbb{E}(N_t^2) & = & \braket{1|a|\Phi_t}.
\end{eqnarray}

The last term that is needed is the product of a coherent and a base state, where the standard form $\braket{z|n} = \bar{z}^n$ is found, which means the probability generating function is just $\braket{z|\Phi_t} = \sum_{n=0}^\infty p_n(t)\bar{z}^n$.

%%%%%%%%%%%%%%%%%%%%%%%%%%%%%%%%%%%%%%%%%%%%%%%%%%%%%%%%%%%%%%%%%%%%%%%%%%%

\subsection{Path Integral Construction}
\label{PIC2}

Now that the machinery has been developed, path integrals of interest can be formed. This relies on a suitable resolution of identity. To this end, the following identity is defined,
\begin{equation}
I = \int \frac{\dd z^2}{\pi}2|z|^\nu K_\nu(2|z|)\ket{z}\bra{z},
\label{Roi2}
\end{equation}
where $K_\nu$ is a modified Bessel function of the second kind \cite{Bowman2012}, and integration is over the entire complex domain. The validy of this will be seen in the next subsection, where properties of the associated reproducing kernel are considered.

To construct a path integral for features of interest, an initial state is also needed. Given the framework above, the most convenient form will be seen to be $\ket{\Phi_0} = \hat{I}_\nu(w)^{-1}\ket{w}$.

Then if we consider the path integral for the generating function, time slicing in the usual fashion results in the expression,
\begin{eqnarray}
\hat{I}_\nu(w)G(x,w;t) & = & \int \mathcal{D}z \hh\braket{x|z_n}\prod_{k=1}^n\braket{z_k|e^{\Delta L(a^\dag,a;t_k)}|z_{k-1}}\braket{z_0|w}\nonumber\\
& = & \int \mathcal{D} z \hh \prod_{k=1}^n(1+\Delta L(\bar{z}_k,z_{k-1};t_k))\prod_{k=0}^{n+1}\hat{I}_\nu(\bar{z}_kz_{k-1}),
\label{PI2}
\end{eqnarray}
where we have measure $\int \mathcal{D}z = \prod_{k=0}^n\int\frac{\ddd z_k^2}{\pi}|z|^\nu K_\nu(2|z_k|)$, and set $z_{n+1}\equiv 1$ and $z_{-1} \equiv w$.

This will later be used for expansion methods and to find some exact results (in \S \ref{TSE2}). First we need some properties of the associated reproducing kernel.

%%%%%%%%%%%%%%%%%%%%%%%%%%%%%%%%%%%%%%%%%%%%%%%%%%%%%%%%%%%%%%%%%%%%%%%%%%%

\subsection{The Reproducing Kernel}
\label{TRK2}

Now given the normalisation properties in Eq. \ref{Eig2} we can pre and post multiply Eq. \ref{Roi2} by $\bra{u}$ and $\ket{v}$ to get the expression,
\begin{equation}
\hat{I}_v(\bar{u}v) = \int\frac{\dd z^2}{\pi}2|z|^\nu K_\nu(2|z|)\hat{I}_\nu(\bar{u}z)\hat{I}_\nu(\bar{z}v).
\label{TR2}
\end{equation}
Thus we have a reproducing kernel $\hat{I}_\nu(\bar{u}v)$ with measure $\dd \mu(z) = \frac{2}{\pi}\dd z^2\hh|z|^\nu K_\nu(2|z|)$. Coherent state formalism of this nature was originally developed for charged bosons \cite{Bhaumik1976, Klauder1985}, although the most comprehensive and transparent exposition can be found in a more recent application to Landau levels \cite{Aremua2015}.

Now from the series expansion for $\hat{I}_\nu$ we find $\Delta_x \hat{I}_\nu(xy) = y\hat{I}_\nu(xy)$, where we define the differential operator $\Delta_x \equiv x^{-\nu}\partial_xx^{1+\nu}\partial_x$. Thus from the expression above we find,
\begin{equation}
\int\dd \mu(z)\hh\bar{z}^mz^n\hat{I}_\nu(\bar{u}z)\hat{I}_\nu(\bar{z}v) = \Delta_v^m\Delta_{\bar{u}}^n\hat{I}_\nu(\bar{u}v).
\end{equation}

Thus for functions $f(z)$ or $g(\bar{z})$ that contain one of $z$ or its conjugate we get the simple reproducing results,
\begin{eqnarray}
\int\dd \mu(z)\hh f(z)\frac{\hat{I}_\nu(\bar{u}z)\hat{I}_\nu(\bar{z}v)}{\hat{I}_\nu(\bar{u}v)} & = & f(v),
\nonumber\\
\int\dd \mu(z)\hh g(\bar{z})\frac{\hat{I}_\nu(\bar{u}z)\hat{I}_\nu(\bar{z}v)}{\hat{I}_\nu(\bar{u}v)} & = & f(\bar{u}).
\label{SR2}
\end{eqnarray}
For functions involving both variables, the terms interact and the result is more complicated. Consider a function $f(\bar{z},z)$ written such that $\bar{z}$ is left of $z$, along with a function $g(z,\bar{z})$ with $z$ followed by $\bar{z}$. Then we find that,
\begin{eqnarray}
\int\dd \mu(z)\hh f(\bar{z},z)\hat{I}_\nu(\bar{u}z)\hat{I}_\nu(\bar{z}v) 
& = & f(\Delta_v, v)\hat{I}_\nu(\bar{u}v).
\nonumber\\
\int\dd \mu(z)\hh g(z,\bar{z})\hat{I}_\nu(\bar{u}z)\hat{I}_\nu(\bar{z}v) 
& = & g(\Delta_{\bar{u}},\bar{u})\hat{I}_\nu(\bar{u}v).
\label{Recurr2}
\end{eqnarray}

This is the form that will be applied to the path integral in Eq. \ref{PI2} (cf. Eq.s \ref{Recurr1a},\ref{Recurr1b}).

%%%%%%%%%%%%%%%%%%%%%%%%%%%%%%%%%%%%%%%%%%%%%%%%%%%%%%%%%%%%%%%%%%%%%%%%%%%

\subsection{Time Series Expansion}
\label{TSE2}

The expansion methods are now very similar to the previous section. First the path integral is calculated, which can again be performed forward or backward in time. Integrating forward or backward in time induces the same ordering requirements as the last section, and we end up with the following form,
\begin{eqnarray}
\hat{I}_v(w)G(x,w;t) & = & \overrightarrow{\mathcal{T}} e^{A(t)w(\Delta_w^2-\Delta_w) + B(t)w(1-\Delta_w)}\hat{I}_\nu(\bar{x}w), 
\label{GFQ1}\\
\hat{I}_v(w)G(x,w;t) & = & \overleftarrow{\mathcal{T}} e^{A(t)(\bar{x}^2-\bar{x})\Delta_{\bar{x}} + B(t)(1-\bar{x})\Delta_{\bar{x}}}\hat{I}_\nu(\bar{x}w), 
\label{GFQ2}
\end{eqnarray}
where we again have integrated rates $A(t) = \int_0^t \dd s\hh \alpha(s)$ and $B(t) = \int_0^t\dd s\hh \beta(s)$. These can now be expanded in much the same way as \S \ref{LBDP}. There the expansion was performed in terms of $w$ dependent operators. However, the expansion can also be done on the generating function variable $x$, which we next exemplify.

%%%%%%%%%%%%%%%%%%%%%%%%%%%%%%%%%%%%%%%%%%%%%%%%%%%%%%%%%%%%%%%%%%%%%%%%%%%

\subsubsection{Pure Birth}

If we take the pure birth case ($\beta(t) = 0$) we find time ordering is not important, and expanding Eq. \ref{GFQ2} (choosing $x$ to be real),
\begin{equation}
\hat{I}_v(w)G(x,w;t) = \sum_{k=0}^\infty\frac{A(t)^k}{k!}(x^2\Delta_x - x\Delta_x)^k
\sum_{m=0}^\infty\frac{(xw)^m}{\Gamma(m+1)\Gamma(m+\nu+1)}.
\end{equation}
Then note that we have the following operator actions, where we use shorthand $(m)_\nu \equiv m(m+\nu)$ (this should not be confused with the Pochhammer symbol, which is not used in this section),
\begin{eqnarray}
x^2\Delta_x: x^m & \longrightarrow & (m)_\nu x^{m+1}, \nonumber \\
-x\Delta_x: x^m & \longrightarrow & -(m)_\nu x^m.
\end{eqnarray}
Thus we have a walk that either goes up one step or moves horizontally, both with the same coefficient $(m)_\nu$ (albeit with different signs). Now if we have a path over $k$ steps that includes $r$ increases, then we must have factors $(m)_\nu,(m+1)_\nu,\dots,(m+r)_\nu$ occurring at the up steps. The remaining $k-r$ horizontal steps are composed from $(-1)^{k-r}$ and a subset of these factors (with repetition allowed). Then we find that if $X_{k,r}$ sums the products across the paths, that (cf. Eq. \ref{Xpaths}),
\begin{equation} 
X_{k,r} = (m)_\nu(m+1)_\nu\dots(m+r)_\nu\sum_{\{\underline{\eta}:\sum_{i=0}^r \eta_i =  k-r\}}(m)_\nu^{\eta_0}(m+1)_\nu^{\eta_1}\dots(m+r)_\nu^{\eta_r}(-1)^{k-r}.
\end{equation}
The generating function can then be written as,
\begin{equation}
G(x,w;t) = I_\nu(w)^{-1}\sum_{k=0}^\infty\frac{A(t)^k}{k!}
\sum_{m=0}^\infty\frac{(xw)^m}{\Gamma(m+1)\Gamma(m+\nu+1)}
\sum_{r=0}^kX_{k,r}x^r.
\end{equation}

Again like \S\ref{LBDP} we have an inner product for the generating function, this time between an exponential and a modified Bessel function. Note that the mixing factor $\sum_{r=0}^kX_{k,r}x^r$ now involves $x$ (rather than $w$) meaning getting moments is slightly more awkward, but the results of expanding with respect to $x$ or $w$ are similar. Note that the expansion in terms of $w$ would involve down steps rather than up steps. 

%%%%%%%%%%%%%%%%%%%%%%%%%%%%%%%%%%%%%%%%%%%%%%%%%%%%%%%%%%%%%%%%%%%%%%%%%%%

\subsubsection{Pure Death}

If we take the pure death case ($\alpha(t) = 0$) we similarly find, expanding Eq. \ref{GFQ2}, that (again choosing $x$ to be real),
\begin{equation}
\hat{I}_v(w)G(x,w;t) = \sum_{k=0}^\infty\frac{B(t)^k}{k!}((1-x)\Delta_x)^k
\sum_{m=0}^\infty\frac{(xw)^m}{\Gamma(m+1)\Gamma(m+\nu+1)}.
\end{equation}

Then the following operator actions are needed, where,
\begin{eqnarray}
\Delta_x: x^m & \longrightarrow & (m)_\nu x^{m-1}, \nonumber \\
-x\Delta_x: x^m & \longrightarrow & -(m)_\nu x^m.
\end{eqnarray}
Then we have the same situation as before, except now we have down steps. If $X_{k,r}$ again sums the factors over paths, we similarly find,
\begin{equation} 
X_{k,r} = (m)_\nu(m-1)_\nu\dots(m-r)_\nu\sum_{\{\underline{\eta}:\sum_{i=0}^r \eta_i =  k-r\}}(m)_\nu^{\eta_0}(m-1)_\nu^{\eta_1}\dots(m-r)_\nu^{\eta_r}(-1)^{k-r}.
\end{equation}
The generating function then takes a similar form to above (albeit with distinct $X_{k,r}$),
\begin{equation}
G(x,w;t) = \hat{I}_\nu(w)^{-1}\sum_{k=0}^\infty\frac{B(t)^k}{k!}
\sum_{m=0}^\infty\frac{(xw)^m}{\Gamma(m+1)\Gamma(m+\nu+1)}
\sum_{r=0}^{\min\{k,m\}}X_{k,r}x^{-r}.
\end{equation} 

%%%%%%%%%%%%%%%%%%%%%%%%%%%%%%%%%%%%%%%%%%%%%%%%%%%%%%%%%%%%%%%%%%%%%%%%%%%

\subsubsection{Time Independent Birth and Death}

When both processes are included we need to consider time ordering. For time independent rates, the path walking approach can be used, giving a solution semi-numerical in nature. We now have all four operators above involved and so the paths can move up, down and horizontally (with two choices of coefficient). If $X_{k,r}$ sums the coefficients over paths, the following recurrence is applicable,
\begin{equation}
X_{k,r} = \alpha(m+r-1)_\nu X_{k-1,r-1} - \alpha(m+r)_\nu X_{k-1,r} +
\beta(m+r+1)_\nu X_{k-1,r+1} - \beta(m+r)_\nu X_{k-1,r}.
\end{equation}
The generating function is then,
\begin{equation}
G(x,w;t) = I_\nu(w)^{-1}\sum_{k=0}^\infty\frac{t^k}{k!}
\sum_{m=0}^\infty\frac{(xw)^m}{\Gamma(m+1)\Gamma(m+\nu+1)}
\sum_{r=-\min\{k,m\}}^{k}X_{k,r}x^r.
\end{equation} 

%%%%%%%%%%%%%%%%%%%%%%%%%%%%%%%%%%%%%%%%%%%%%%%%%%%%%%%%%%%%%%%%%%%%%%%%%%%

\subsubsection{Time Dependent Birth and Death}

For time dependent rates $\alpha(t)$ and $\beta(t)$ the time ordering needs to be considered in much the same way as \S \ref{LBDP} giving a numerical, semi-analytic approach. This requires a bit of formalism and it is a bit simpler to consider operators acting on $w$ rather than $x$. We introduce a generalised modified, modified Bessel function,
\begin{equation}
\hat{I}_\nu^{r}(w;f) = \sum_{m=\max\{0,-r\}}^\infty \frac{w^{m+r}f(m)}{\Gamma(m+1)\Gamma(m+\nu+1)}.
\end{equation}

Now, from Eq. \ref{GFQ1} we can write the expected size as the following expansion (higher order moments can be obtained in much the same way),
\begin{equation}
\hat{I}_\nu(w)\mathbb{E}(N_t) = \overrightarrow{\mathcal{T}} \sum_{k=0}^\infty\frac{(A(t)\mathcal{A}+B(t)\mathcal{B})^k}{k!}I_\nu^0(w;m),
\end{equation}
where we have operators $\mathcal{A} = w(\Delta_w^2-\Delta_w)$ and $\mathcal{B} = w(1-\Delta_w)$. Now we know that $\Delta_w:w^m \longrightarrow (m)_\nu w^{m-1}$, and so we find actions,
\begin{eqnarray}
\mathcal{A}\cdot I_\nu^r(w;f) & = & I_\nu^{r-1}(w;(m+r)_\nu^2f)-I_\nu^r(w;(m+r)_\nu f), \nonumber\\
\mathcal{B}\cdot I_\nu^r(w;f) & = & I_\nu^{r+1}(w;f)-I_\nu^r(w;(m+r)_\nu f). 
\end{eqnarray}

Then, the zeroth order term ($k=0$) for the mean requires no time ordering, and we get $I_\nu(w)^{-1}I_\nu^0(w;m)$. The first order term ($k=1$) also requires no time ordering and is found from the action of $(A(t)\mathcal{A}+B(t)\mathcal{B})$ on $I_\nu^0(w;m)$ (details are left to the reader). Second order (and higher) terms require time ordering and is much like \S \ref{LBDP}. For example the action of $\mathcal{B}\mathcal{A}$ arising in the expansion of second order terms ($k=2$) is given by,
\begin{eqnarray}
&& I_\nu(w)^{-1}\int_{0<s<s'<t} \dd s \hh \dd s' \hh \alpha(s')\beta(s) \mathcal{B}\mathcal{A}I_\nu^0(w;m) \\ 
&& = I_\nu(w)^{-1}\left(2I_\nu^0(w;(m)_\nu^2m) -I_\nu^{-1}(w;(m-1)_\nu(m)_\nu^2m) -I_\nu^1(w;(m)_\nu m)\right)\int_{0<s<s'<t} \dd s \hh \dd s' \hh \alpha(s')\beta(s).
\nonumber
\end{eqnarray}
Higher order terms can be obtained similarly, allowing approximations to moments to be evaluated.

%%%%%%%%%%%%%%%%%%%%%%%%%%%%%%%%%%%%%%%%%%%%%%%%%%%%%%%%%%%%%%%%%%%%%%%%%%%
%%%%%%%%%%%%%%%%%%%%%%%%%%%%%%%%%%%%%%%%%%%%%%%%%%%%%%%%%%%%%%%%%%%%%%%%%%%

\section{Diffusion Equations}
\label{DE}

Stochastic PDEs provide a way of modelling Brownian motion and generalizations thereof \cite{Klebaner2012}. Such systems have been put into path integral form previously \cite{Kleinert2009, Onsager1953b, Martin1973, Onsager1953a, Wio2013}. More recently, the work of \cite{Greenman2020, Ohkubo2013} has seen application of the Doi framework to establish duality between diffusion processes and birth-death processes. This section develops a Doi algebra based path integral formalism for diffusion processes that allows time series expansion to be successfully applied, as well as obtaining exact results.

%%%%%%%%%%%%%%%%%%%%%%%%%%%%%%%%%%%%%%%%%%%%%%%%%%%%%%%%%%%%%%%%%%%%%%%%%%%%%%%%%%

\subsection{Setup}
\label{S}

Now, as noted in \cite{Ohkubo2013}, diffusion equations can be modelled in the following way. Take a stochastic PDE of the following form,
\begin{equation}
\dd X_t = \mu(X_t,t) \dd t + \sigma(X_t,t) \dd W_t,
\label{Spde}
\end{equation}
where $\mu(X_t,t)$ is a drift term and $\sigma(X_t,t)$ is a noise term, weighting the standard Brownian motion, $\dd W_t$. Let $P(x,t)$ denote the associated probability density function, which satisfies the Fokker-Planck equation,
\begin{equation}
\frac{\dd P}{ \dd t} = -\mu(x,t) \frac{\dd P}{\dd x} + \frac{1}{2}\sigma(x,t)^2\frac{\dd^2 P}{\dd x^2}.
\label{FP}
\end{equation}
We assume for simplicity that the process starts at a known fixed value $X_0=w$. 

Now, to use Doi formalism, the following state representation is adopted,
\begin{equation}
\ket{\Phi_t} = \int \dd x \hh P(x,t) \ket{x},
\label{CtsRep}
\end{equation}
where the states $\ket{x}$ are coherent (in the sense of \S \ref{LBDP}). The initial condition is thus simply represented by the state $\ket{\Phi_0}=\ket{w}$, and the Fokker-Planck equation can be re-written as,
\begin{equation} 
\frac{\partial}{\partial t}\ket{\Phi_t} = L\ket{\Phi_t},
\end{equation}
where the Liouvillian operator $L$ is given by,

\begin{equation}
L(a^\dag,a;t)= a^\dag\mu(a,t) +\frac{1}{2}(a^\dag)^2\sigma(a,t)^2.
\label{DiffL}
\end{equation}

The sign change between Eq.s \ref{FP} and \ref{DiffL} is due to an integration by parts when using the representation in Eq. \ref{CtsRep}. Note also that the order of operators $a$ and $a^\dag$ representing $x$ and $\frac{\partial}{\partial x}$ gets swapped. This formalism also relies on the assumption that $\mu$ and $\sigma^2$ are polynomial functions of $x$. Note that the operators are then in natural order. 

Now to construct features of interest note that the coherent state $\bra{0}$ (which is also the null state) plays the role of the sum operator usually seen in Doi formalism. We find that probability conservation is given by the expression $L\ket{\Phi_t} = 0$. It can also be seen that $\braket{iy|\Phi_t}$ is the characteristic function, and $\braket{s|\Phi_t}$ is the moment generating function. We shall consider correlation functions, such as the $m^\mathrm{th}$ factorial moment $\mathbb{E}_X((X_t)^m) = \braket{0|a^m|\Phi_t}$. To do this, we next we build associated path integrals. 

%%%%%%%%%%%%%%%%%%%%%%%%%%%%%%%%%%%%%%%%%%%%%%%%%%%%%%%%%%%%%%%%%%%%%%%%%%%

\subsection{Exact Calculation}
\label{EC}

To construct a path integral requires the construction of a resolution of identity. We could use that of Eq. \ref{ROI1}, however, the following form offers simpler exact calculations and offers an alternative reproducing kernel to consider, 
\begin{equation}
I = \iint \frac{\dd u\hh \dd v}{2\pi}e^{-iuv}\ket{iv}\bra{u}.
\label{ROI3}
\end{equation}
Here $u$ and $v$ are real variables \cite{Greenman2017}. Then time slicing results in the following matrix element, 
\begin{eqnarray}
\braket{z|e^{\int_0^t \ddd s \hh L(a^\dag,a;s)}|w} & = & \iint \prod_{k=0}^n\frac{\dd u_k\hh \dd v_k}{2\pi}
\exp\left\{
-i\sum_{k=0}^nu_kv_k+i\sum_{k=1}^nu_kv_{k-1}\right.\nonumber\\
&& \hspace{3cm} \left.+izv_n+\sum_{k=1}^N\Delta L(u_k,iv_{k-1};t_k)+u_0w
\right\}.
\label{EPI3}
\end{eqnarray}

This is the form that we shall use for time series expansion. However, to see the path integral continuum approach, consider the example with $\mu(a,t) = \alpha(t)a +\beta(t)$ and $\sigma(a,t)^2 = \sigma(t)^2$, that is, scaled, time dependent Brownian motion with linear drift. Then taking the continuum limit, and forming a generating functional by adding a source term with function $J$ to the resultant action, gives the following (an integration by parts also takes place in the action),
\begin{eqnarray}
G(z,x,J;t) & = & \iint \mathcal{D}u\mathcal{D}v\exp\left\{
iv(t)(z-u(t))+u(0)x+\int_0^t \dd s\hh\left[\beta(s) u(s) +\frac{1}{2}\sigma(s)^2u(s)^2\right]
\right.\nonumber\\
&&\left.+\int_0^t \dd s\hh iv(s)\left[\frac{\partial u(s)}{\partial s}+\alpha(s) u(s) +J(s)\right]
\right\},
\end{eqnarray}
where $\displaystyle\iint \mathcal{D}u\mathcal{D}v = \lim_{n \rightarrow \infty}\iint\prod_{k=0}^n\frac{\dd u_k\hh \dd v_k}{2\pi}$.

Then integration over the $v$ variable forces the conditions,
\begin{eqnarray}
u(t) & = & z,\nonumber\\
\frac{\partial u(s)}{\partial s} & = & -\alpha(s) u(s) -J(s).
\end{eqnarray}
Integration over the $u$ variables then results in the solution,
\begin{eqnarray}
G(z,x,J;t) & = & \exp\left\{u(0)x+\int_0^t \dd s\hh[\beta(s) u(s) +\frac{1}{2}\sigma(s)^2u(s)^2]
\right\},\nonumber\\
u(s) & = & z\exp\left\{\int_s^t \dd s'\hh\alpha(s')\right\}+\int_s^t \dd s' \hh J(s') \exp\left\{\int_{s}^{s'}\dd s''\hh\alpha(s'')\right\}.
\end{eqnarray}

This allows us to get correlation functions of interest. For example, the mean is given by the functional derivative,

\begin{equation}
\mathbb{E}(X_t) = \left.\frac{\delta G(0,x,J;t)}{\delta J(t)}
\right|_{J \equiv 0} = x\exp\left\{\int_0^t \dd s\hh\alpha(s)\right\} + \int_0^t \dd s \hh \beta(s)\exp\left\{ \int_s^t \dd s'\hh \alpha(s')\right\},
\end{equation} 
which can be quickly found directly from the stochastic PDE in Eq. \ref{Spde}. However, higher order moments are less easy to get this way, but can be readily obtained from the generating functional. Other features of interest, such as the characteristic function $\mathbb{E}_X(e^{iyX_t})=G(iy,x,0,t)$ are also available from the same formalism.

Note that this path integral, obtained with the resolution of identity in Eq. \ref{ROI3}, gives the same form of action that a Bargman-Fock type approach used by Peliti \cite{Peliti1985} would produce, although the derivation is slightly different. 

%%%%%%%%%%%%%%%%%%%%%%%%%%%%%%%%%%%%%%%%%%%%%%%%%%%%%%%%%%%%%%%%%%%%%%%%%%%

\subsection{Expansion Methods}
\label{EM}

Expansion methods can also be applied to these systems, as is now demonstrated. To highly them, we analyse the characteristic function which can be written as $\mathbb{E}(e^{iyX_t}) = \braket{iy|\overleftarrow{\mathcal{T}}e^{\int_0^t \ddd s\hh L(a^\dag,a;s)}|w}$. To implement the expansion methods, we need to identify the an associated reproducing kernel system. Now, if we left and right multiply Eq. \ref{ROI3} by $\bra{x}$ and $\ket{iy}$, respectively, note that,
\begin{equation}
e^{ixy} = \int \frac{\dd u\hh\dd v}{2\pi}e^{-iuv}e^{ixv}e^{iuy} = \int \dd u \hh\delta(u-x)e^{iuy} = \int \dd v \hh\delta(v-y)e^{ixv}.
\end{equation}
This is an integral over delta functions and (unlike Eq. \ref{KK1}) is interpreted in the distribution sense. Note that there is some latitude in this identity. If $x$ (resp. $y$) is real, then $y$ (resp. $x$) can be any complex number.  Now from this we obtain,
\begin{equation}
\int \frac{\dd u\hh\dd v}{2\pi}e^{-iuv}u^m(iv)^ne^{ixv}e^{iuy}
= \partial_{iy}^m\partial_x^ne^{ixy}.
\label{RK3}
\end{equation}
This results in the more general forms,
\begin{eqnarray}
\int \frac{\dd u \hh \dd v}{2\pi} e^{-iuv}f(u,iv)e^{ixv} e^{iuy} & = & f(\partial_{iy},iy)e^{ixy},
\nonumber\\
\int \frac{\dd u \hh \dd v}{2\pi} e^{-iuv}g(iv,u)e^{ixv} e^{iuy} & = & g(\partial_{x},x)e^{ixy}.
\label{Recurr3}
\end{eqnarray}
Note that the transformation of the function $f$ is essentially equivalent to that of Eq.s \ref{Recurr1a} and \ref{Recurr1b}. Using the associated kernel properties of \S \ref{LBDP} could have been used in this section to give the same results. 

Then, from Eq. \ref{EPI3}, the path integral for the characteristic function, for example, can be expressed as,
\begin{equation}
\mathbb{E}(e^{iyX_t}) = \braket{iy|e^{\int_0^t \ddd s \hh L(a^\dag,a;s)}|w} = \iint \mathcal{D}u\hh\mathcal{D}v
\exp\left\{
-i\sum_{k=0}^nu_kv_k+i\sum_{k=0}^{n+1}u_kv_{k-1}\right\}
\prod_{k=1}^n(1+\Delta L(u_k,iv_{k-1};t_k)),
\end{equation}
where we have $u_{n+1} \equiv 0$ and $v_{-1} \equiv -iw$, and measure $\iint\mathcal{D}u\mathcal{D}v\equiv\int \prod_{k=0}^n\frac{\ddd u_k\hh \ddd v_k}{2\pi}$.

Then doing the integration term by term gives the following, where $M(w,t)=\int_0^t \dd s \hh\mu(w,s)$ and $S(w,t)^2=\int_0^t \dd s \hh\sigma^2(w,s)$ are the cumulative drift and noise,
\begin{equation}
\mathbb{E}(e^{iyX_t}) = e^{M(w,t)\partial_w+\frac{1}{2}S(w,t)^2\partial_w^2}e^{iyw}.
\end{equation}

The expansion techniques are then similar to previous sections. If we consider geometric Brownian motion as a simple example, we have drift $\mu(X_t,t) = \mu X_t$ and noise $\sigma(X_t,t) = \sigma X_t$ terms, we find,
\begin{equation}
\mathbb{E}(e^{iyX_t}) = \sum_{k=0}^\infty \frac{t^k}{k!}(\mu w\partial_w+\frac{1}{2}\sigma^2w^2\partial_w^2)^k
\sum_{m=0}^\infty \frac{(iyw)^m}{m!}.
\end{equation}
Then the operators to consider are,
\begin{eqnarray}
w\partial_w: w^m & \longrightarrow & m w^{m}, \nonumber \\
w^2\partial_w^2: w^m & \longrightarrow & (m)_2 w^m.
\end{eqnarray}
Thus we have neither up or down steps to consider, just horizontal steps. This makes the path counting exercise simple and we find,
\begin{equation}
\mathbb{E}(e^{iyX_t}) = \sum_{m=0}^\infty\sum_{k=0}^\infty \frac{t^k}{k!}(\mu m+\frac{1}{2}\sigma^2m(m-1))^k \frac{(iyw)^m}{m!} = 
\sum_{m=0}^\infty e^{t(\mu m+\frac{1}{2}\sigma^2m(m-1))} \frac{(iyw)^m}{m!}.
\end{equation}

This gives the correct characteristic function, from which the $m^\mathrm{th}$ moment $w^me^{t(\mu m+\frac{1}{2}\sigma^2m(m-1))}$ can be read off. This result can be derived by classical methods. However, for more complicated diffusions, the path walking approach offers an alternative method of investigation.

%%%%%%%%%%%%%%%%%%%%%%%%%%%%%%%%%%%%%%%%%%%%%%%%%%%%%%%%%%%%%%%%%%%%%%%%%%%
%%%%%%%%%%%%%%%%%%%%%%%%%%%%%%%%%%%%%%%%%%%%%%%%%%%%%%%%%%%%%%%%%%%%%%%%%%%

\section{Conclusions}
\label{ConcS}

We have introduced a time series expansion method for path integrals across a range of stochastic models including birth-death and diffusion processes. This results in series solutions for features of interest such as correlation, characteristic and moment generating functions. For processes with a single time dependent rate function, these series can often be fully expressed in entirety. For processes with more than one time dependent rate, this is not particularly trivial due to the time ordering effects. However, it still provides a method of approximation such that approximating terms have closed (rather than just numerical) form. 

This method is somewhat akin to the Dyson expansion methods of quantum mechanics, except that instead of splitting the Hamiltonian into a standard and complex part, the series is calculated directly with the aid of an underlying reproducing kernel space. This has the advantage for stochastic processes where the Liouvillian (that is, the Hamiltonian for the stochastic process) cannot in general be separated into time dependent and independent operator parts, requisite for Dyson expansion methods.

The process is semi-analytic, and can be implemented numerically with relative simplicity. When the series cannot be explicitly evaluated in its entirety and is instead considered as a numerical method of approximation, there are some computational limitations. These arise either because greater numerical precision is required, or the number of terms in the expansion gets too large, causing either memory or run time constraint problems. 

The algebra used for standard birth-death processes is just that of Doi-Peliti methods. However we have shown that $\mathfrak{su}(1,1)$ can be used to model birth-death processes with quadratic rates. This does open the possibility that other algebras (Lie or otherwise) may prove useful when analysing other systems of interest with these kind of techniques.

%%%%%%%%%%%%%%%%%%%%%%%%%%%%%%%%%%%%%%%%%%%%%%%%%%%%%%%%%%%%%%%%%%%%%%%%%%%
%%%%%%%%%%%%%%%%%%%%%%%%%%%%%%%%%%%%%%%%%%%%%%%%%%%%%%%%%%%%%%%%%%%%%%%%%%%

{\scriptsize
\bibliographystyle{abbrv}
\bibliography{refs_TimeSeries}

\begin{thebibliography}{10}

\bibitem{Aremua2015}
I.~Aremua, M.~N. Hounkonnou, and E.~Balo{\"\i}tcha.
\newblock Coherent states for landau levels: algebraic and thermodynamical
  properties.
\newblock {\em Reports on Mathematical Physics}, 76(2):247--269, 2015.

\bibitem{Bartlett1978}
M.~S. Bartlett.
\newblock {\em An introduction to stochastic processes: with special reference
  to methods and applications}.
\newblock CUP Archive, 1978.

\bibitem{Beals2016}
R.~Beals and R.~Wong.
\newblock {\em Special functions and orthogonal polynomials}, volume 153.
\newblock Cambridge University Press, 2016.

\bibitem{Bhaumik1976}
D.~Bhaumik, K.~Bhamik, and B.~Dutta-Roy.
\newblock Charged bosons and the coherent state.
\newblock {\em Journal of Physics A: Mathematical and General},
  9(9):1507--1512, 1976.

\bibitem{Bogoliubov1946}
N.~Bogoliubov.
\newblock Kinetic equations.
\newblock {\em Journal of Physics USSR}, 10(3):265--274, 1946.

\bibitem{Born1949}
M.~Born and H.~S. Green.
\newblock {\em A general kinetic theory of liquids}.
\newblock CUP Archive, 1949.

\bibitem{Bowman2012}
F.~Bowman.
\newblock {\em Introduction to Bessel functions}.
\newblock Courier Corporation, 2012.

\bibitem{Buice2007}
M.~A. Buice and J.~D. Cowan.
\newblock Field-theoretic approach to fluctuation effects in neural networks.
\newblock {\em Physical Review E}, 75(5):051919, 2007.

\bibitem{Cardy1998}
J.~L. Cardy and U.~C. T{\"a}uber.
\newblock Field theory of branching and annihilating random walks.
\newblock {\em Journal of statistical physics}, 90(1):1--56, 1998.

\bibitem{Greenman2016}
T.~Chou and C.~D. Greenman.
\newblock A hierarchical kinetic theory of birth, death and fission in
  age-structured interacting populations.
\newblock {\em Journal of statistical physics}, 164(1):49--76, 2016.

\bibitem{Daniell1919}
P.~J. Daniell.
\newblock Integrals in an infinite number of dimensions.
\newblock {\em Annals of Mathematics}, pages 281--288, 1919.

\bibitem{Dobramysl2018}
U.~Dobramysl, M.~Mobilia, M.~Pleimling, and U.~C. T{\"a}uber.
\newblock Stochastic population dynamics in spatially extended predator--prey
  systems.
\newblock {\em Journal of Physics A: Mathematical and Theoretical},
  51(6):063001, 2018.

\bibitem{Doi1976a}
M.~Doi.
\newblock Second quantization representation for classical many-particle
  system.
\newblock {\em Journal of Physics A: Mathematical and General}, 9(9):1465,
  1976.

\bibitem{Doi1976b}
M.~Doi.
\newblock Stochastic theory of diffusion-controlled reaction.
\newblock {\em Journal of Physics A: Mathematical and General}, 9(9):1479,
  1976.

\bibitem{Engel2001}
K.-J. Engel and R.~Nagel.
\newblock One-parameter semigroups for linear evolution equations.
\newblock In {\em Semigroup forum}, volume~63, pages 278--280. Springer, 2001.

\bibitem{Feller1939}
W.~Feller.
\newblock Die grundlagen der volterraschen theorie des kampfes ums dasein in
  wahrscheinlichkeitstheoretischer behandlung.
\newblock {\em Acta biotheoretica}, 5(1):11--40, 1939.

\bibitem{Feynman2010}
R.~P. Feynman, A.~R. Hibbs, and D.~F. Styer.
\newblock {\em Quantum mechanics and path integrals}.
\newblock Courier Corporation, 2010.

\bibitem{Garrahan2007}
J.~P. Garrahan, R.~L. Jack, V.~Lecomte, E.~Pitard, K.~van Duijvendijk, and
  F.~van Wijland.
\newblock Dynamical first-order phase transition in kinetically constrained
  models of glasses.
\newblock {\em Physical review letters}, 98(19):195702, 2007.

\bibitem{Gilmore2012}
R.~Gilmore.
\newblock {\em Lie groups, Lie algebras, and some of their applications}.
\newblock Courier Corporation, 2012.

\bibitem{Greenman2017}
C.~D. Greenman.
\newblock A path integral approach to age dependent branching processes.
\newblock {\em Journal of Statistical Mechanics: Theory and Experiment},
  2017(3):033101, 2017.

\bibitem{Greenman2018}
C.~D. Greenman.
\newblock Doi--peliti path integral methods for stochastic systems with partial
  exclusion.
\newblock {\em Physica A: Statistical Mechanics and its Applications},
  505:211--221, 2018.

\bibitem{Greenman2020}
C.~D. Greenman.
\newblock Duality relations between spatial birth--death processes and
  diffusions in hilbert space.
\newblock {\em Journal of Physics A: Mathematical and Theoretical},
  53(44):445002, 2020.

\bibitem{Greenman2015}
C.~D. Greenman and T.~Chou.
\newblock Kinetic theory of age-structured stochastic birth-death processes.
\newblock {\em Physical Review E}, 93:012112, 2015.

\bibitem{Itzykson2012}
C.~Itzykson and J.-B. Zuber.
\newblock {\em Quantum field theory}.
\newblock Courier Corporation, 2012.

\bibitem{Jarvis2005}
P.~D. Jarvis, J.~Bashford, and J.~Sumner.
\newblock Path integral formulation and feynman rules for phylogenetic
  branching models.
\newblock {\em Journal of Physics A: Mathematical and General}, 38(44):9621,
  2005.

\bibitem{Karlin1957b}
S.~Karlin and J.~McGregor.
\newblock The classification of birth and death processes.
\newblock {\em Transactions of the American Mathematical Society},
  86(2):366--400, 1957.

\bibitem{Karlin1957a}
S.~Karlin and J.~L. McGregor.
\newblock The differential equations of birth-and-death processes, and the
  stieltjes moment problem.
\newblock {\em Transactions of the American Mathematical Society},
  85(2):489--546, 1957.

\bibitem{Kendall1948}
D.~G. Kendall.
\newblock {On the Generalized ``{Birth-and-Death}" Process}.
\newblock {\em Ann. Math. Statist.}, 19:1--15, 1948.

\bibitem{Kirkwood1946}
J.~G. Kirkwood.
\newblock The statistical mechanical theory of transport processes i. general
  theory.
\newblock {\em The Journal of Chemical Physics}, 14(3):180--201, 1946.

\bibitem{Kirkwood1947}
J.~G. Kirkwood.
\newblock The statistical mechanical theory of transport processes ii.
  transport in gases.
\newblock {\em The Journal of Chemical Physics}, 15(1):72--76, 1947.

\bibitem{Klauder1985}
J.~R. Klauder and B.-S. Skagerstam.
\newblock {\em Coherent states: applications in physics and mathematical
  physics}.
\newblock World scientific, 1985.

\bibitem{Klebaner2012}
F.~C. Klebaner.
\newblock {\em Introduction to stochastic calculus with applications}.
\newblock World Scientific Publishing Company, 2012.

\bibitem{Kleinert2009}
H.~Kleinert.
\newblock {\em Path integrals in quantum mechanics, statistics, polymer
  physics, and financial markets}.
\newblock World scientific, 2009.

\bibitem{Onsager1953b}
S.~Machlup and L.~Onsager.
\newblock Fluctuations and irreversible process. ii. systems with kinetic
  energy.
\newblock {\em Physical Review}, 91(6):1512, 1953.

\bibitem{Martin1973}
P.~C. Martin, E.~Siggia, and H.~Rose.
\newblock Statistical dynamics of classical systems.
\newblock {\em Physical Review A}, 8(1):423, 1973.

\bibitem{Mcquarrie1964}
D.~A. McQuarrie, C.~Jachimowski, and M.~Russell.
\newblock Kinetics of small systems. ii.
\newblock {\em The Journal of Chemical Physics}, 40(10):2914--2921, 1964.

\bibitem{Ohkubo2013b}
J.~Ohkubo.
\newblock Algebraic probability, classical stochastic processes, and counting
  statistics.
\newblock {\em Journal of the Physical Society of Japan}, 82(8):084001, 2013.

\bibitem{Ohkubo2013}
J.~Ohkubo.
\newblock Extended duality relations between birth--death processes and partial
  differential equations.
\newblock {\em Journal of Physics A: Mathematical and Theoretical},
  46(37):375004, 2013.

\bibitem{Onsager1953a}
L.~Onsager and S.~Machlup.
\newblock Fluctuations and irreversible processes.
\newblock {\em Physical Review}, 91(6):1505, 1953.

\bibitem{Peliti1985}
L.~Peliti.
\newblock Path integral approach to birth-death processes on a lattice.
\newblock {\em Journal de Physique}, 46(9):1469--1483, 1985.

\bibitem{Peskin2018}
M.~Peskin and D.~Scroeder.
\newblock {\em An introduction to quantum field theory}.
\newblock CRC press, 2018.

\bibitem{Rohwer2015}
C.~M. Rohwer and K.~K. M{\"u}ller-Nedebock.
\newblock Operator formalism for topology-conserving crossing dynamics in
  planar knot diagrams.
\newblock {\em Journal of Statistical Physics}, 159(1):120--157, 2015.

\bibitem{Sasaki2009}
R.~Sasaki.
\newblock Exactly solvable birth and death processes.
\newblock {\em Journal of mathematical physics}, 50(10):103509, 2009.

\bibitem{Sattinger2013}
D.~H. Sattinger and O.~L. Weaver.
\newblock {\em Lie groups and algebras with applications to physics, geometry,
  and mechanics}, volume~61.
\newblock Springer Science \& Business Media, 2013.

\bibitem{Schulz2005}
M.~Schulz and P.~Reineker.
\newblock Exact substitute processes for diffusion--reaction systems with local
  complete exclusion rules.
\newblock {\em New Journal of Physics}, 7(1):31, 2005.

\bibitem{Schulz1996}
M.~Schulz and S.~Trimper.
\newblock Parafermi statistics and p-state models.
\newblock {\em Physics Letters A}, 216(6):235--239, 1996.

\bibitem{Tauber2014}
U.~C. T{\"a}uber.
\newblock {\em Critical dynamics: a field theory approach to equilibrium and
  non-equilibrium scaling behavior}.
\newblock Cambridge University Press, 2014.

\bibitem{Tauber2005}
U.~C. T{\"a}uber, M.~Howard, and B.~P. Vollmayr-Lee.
\newblock Applications of field-theoretic renormalization group methods to
  reaction--diffusion problems.
\newblock {\em Journal of Physics A: Mathematical and General}, 38(17):R79,
  2005.

\bibitem{Valent1996}
G.~Valent.
\newblock Exact solutions of some quadratic and quartic birth and death
  processes and related orthogonal polynomials.
\newblock {\em Journal of computational and applied mathematics},
  67(1):103--127, 1996.

\bibitem{VanKampen1992}
N.~G. Van~Kampen.
\newblock {\em Stochastic processes in physics and chemistry}, volume~1.
\newblock Elsevier, 1992.

\bibitem{Wiener1921}
N.~Wiener.
\newblock The average of an analytic functional and the brownian movement.
\newblock {\em Proceedings of the National Academy of Sciences of the United
  States of America}, 7(10):294, 1921.

\bibitem{Wiener1923}
N.~Wiener.
\newblock Differential-space.
\newblock {\em Journal of Mathematics and Physics}, 2(1-4):131--174, 1923.

\bibitem{Wio2013}
H.~S. Wio.
\newblock {\em Path integrals for stochastic processes: An introduction}.
\newblock World Scientific, 2013.

\bibitem{Yvon1935}
J.~Yvon.
\newblock {\em La th{\'e}orie statistique des fluides et l'{\'e}quation
  d'{\'e}tat}, volume 203.
\newblock Hermann \& cie, 1935.

\end{thebibliography}
}

\end{document}